# Synchronous Behavior of Coupled Systems with Discrete Time

A. A. Koronovskiĭ\*, A. E. Hramov, and A. E. Khramova

*Faculty of Nonlinear Processes, Saratov State University, Saratov, 410012 Russia*
*\* e-mail: alkor@cas.ssu.runnet.ru*
Received May 25, 2005

The dynamics of one-way coupled systems with discrete time is considered. The behavior of the coupled logistic maps is compared to the dynamics of maps obtained using the Poincaré sectioning procedure applied to the coupled continuous-time systems in the phase synchronization regime. The behavior (previously considered as asynchronous) of the coupled maps that appears when the complete synchronization regime is broken as the coupling parameter decreases, corresponds to the phase synchronization of flow systems, and should be considered as a synchronous regime. A quantitative measure of the degree of synchronism for the interacting systems with discrete time is proposed. © *2005 Pleiades Publishing, Inc.*

The process of chaotic synchronization of dynamical systems is among the basic nonlinear phenomena extensively studied in recent years, and this process is also of considerable practical significance. The chaotic synchronization can be considered in systems with both discrete time (maps) and continuous time (flows). Although the two classes of dynamical systems are closely interrelated (it is well known that the systems with continuous time can be reduced to maps using the Poincaré sectioning procedure), there are substantial differences between such systems, which accounts for the fact that the phenomena of synchronization of flows and maps are described using different terms and notions [1].

Evidently, various types of synchronous behavior observed in flows and maps must be interrelated. The aim of this study was to compare the behavior of coupled chaotic systems with continuous time to the dynamics of coupled maps. In particular, it will be demonstrated that the regime of oscillations in coupled maps, which has been considered until recently as asynchronous [2], in fact exhibits signs of synchronism and corresponds to the phase synchronization regime in flow systems [1, 3].

As is known, two coupled identical systems with discrete time at a sufficiently large coupling parameter are featuring the regime of complete (identical) synchronization, whereby the states of these systems coincide. The influence of non-identical features on the regime of complete synchronization was considered in [4]. As the coupling parameter decreases, the saddle orbits built in the synchronous attractor of coupled systems lose stability in the transverse direction, and eventually the regime of complete synchronization (or lag synchronization in flow systems) breaks [2, 5–7] and the chaotic attractor also loses stability in the transverse direction.

It should be noted that analogous phenomena have also been observed in coupled flow systems [8, 9]. In the case of interaction between flow systems with slightly different parameters, the regime of complete synchronization is not established and, instead, the process of lag synchronization takes place. Evidently, by applying the procedure of Poincaré sectioning, it is possible to pass from flow systems to discrete maps. Accordingly, the obtained maps will exhibit complete synchronization. Thus, the complete and lag synchronization in flow systems refer to essentially the same type of synchronous behavior, which is consistent with our recent results [10, 11]. For this reason, below we will not distinguish the regimes of complete synchronization and lag synchronization in flow systems and will use the general term "complete synchronization."

As the coupling parameter decreases, the saddle orbits of coupled flow systems (like those of maps) lose stability in the transverse direction and eventually the regime of complete synchronization (or lag synchronization) in these flows breaks [8, 9]. In the maps obtained using the Poincaré sections, the destruction of synchronism proceeds exactly as described above.

Thus, up to the time when the complete synchronization (including the lag synchronization in flow systems) is broken, the synchronous regimes in flows and maps are completely analogous. It should also be noted that, in both flows [12] and maps [5], the destruction of the complete synchronization regime is accompanied by intermittency of the "on–off" type, which is characterized by the corresponding power laws with equal exponents for the systems with continuous and discrete time.

When the regime of complete synchronization between systems with continuous time is broken (as a result of a decrease in the coupling parameter), the sys-tems pass to a regime of phase synchronization [12], which is a particular case of the time scale synchronization (whereby one part of the time scale is synchronized and the other is not) [10, 11]. At the same time, it is commonly accepted that the destruction of complete synchronization in maps is followed by the onset of asynchronous oscillations. Taking into account that, as was pointed out above, it is always possible to pass from flows to maps with the aid of the Poincaré section, we can pose the following question: "If the systems with continuous time exhibit phase synchronization, why is no such synchronous behavior observed for the maps obtained from these flows using the procedure of Poincaré sectioning?" It is traditionally assumed that this procedure, which reduces the flow systems to maps, excludes numerous states of the system from the consideration (leaving only the states belonging to the surface of the section), so that the residual data are insufficient to recognize the synchronism, and the map dynamics exhibits asynchronous character.

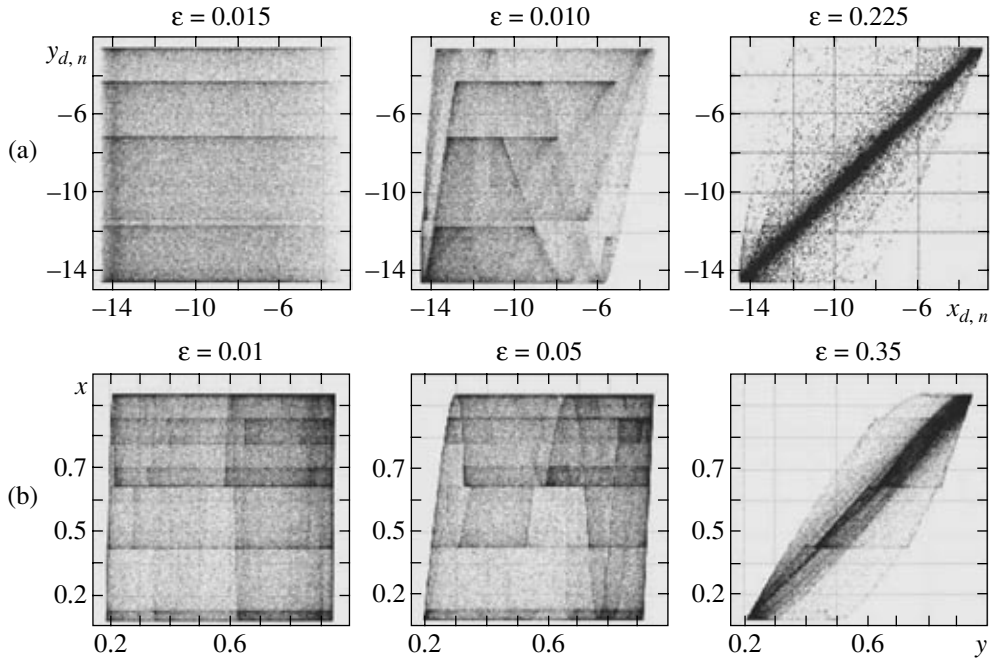

**Fig. 1.** Behavior of one-way coupled chaotic systems: (a) the states $(x_{d,n}, x_{r,n})$ of maps (3) obtained from the Rössler flows (1) by means of the Poincaré sectioning; for the coupling parameter $\varepsilon = 0.015$, the synchronization is absent; for $\varepsilon = 0.10$ and $0.225$, the systems exhibit phase synchronization; (b) the states $(x, y)$ of logistic maps (2) at various values of the coupling parameter $\varepsilon$.

However, as will be shown below, the maps obtained using the Poincaré sections for flow systems occurring in the regime of phase synchronization still carry an "imprint" of the synchronous dynamics and, hence, their behavior has to be considered as synchronized. Moreover, since other discrete maps (e.g., logistic) exhibit generally the same behavior as the maps obtained using the Poincaré sections from flow systems, their dynamics (previously considered as asynchronous) exhibits the features of synchronous behavior and should be considered as synchronized.

The above consideration will be illustrated using a model representing one-way coupled Rössler systems with slightly mismatched parameters:

$$\dot{x}_d = -\omega_d y_d - z_d, \quad \dot{x}_r = -\omega_r y_r - z_r + \varepsilon(x_d - x_r),$$
$$\dot{y}_d = \omega_d x_d + a y_d, \quad \dot{y}_r = \omega_r x_r + a y_r, \qquad (1)$$
$$\dot{z}_d = p + z_d(x_d - c), \quad \dot{z}_r = p + z_r(x_r - c),$$

where $\varepsilon$ is the coupling parameter. The values of the control parameters are selected by analogy with those used in [13]: $a = 0.15$, $p = 0.2$, $c = 10.0$, $\omega_r = 0.95$, $\omega_d = 0.93$. We also consider the logistic maps

$$x_{n+1} = f(x_n, \lambda_x),$$
$$y_{n+1} = f(y_n, \lambda_y) + \varepsilon(f(x_n, \lambda_x) - f(y_n, \lambda_y)), \qquad (2)$$

where $\varepsilon$ is also the coupling parameter, $f(x, \lambda) = \lambda x(1 - x)$, $\lambda_x = 3.75$, and $\lambda_y = 3.79$. Using sections of the phase flow by the Poincaré surfaces $y_d = 0$, $\dot{y}_d < 0$ and $y_r = 0$, $\dot{y}_r < 0$, the drive and response systems with continuous time (1) were reduced to the following coupled two-dimensional maps:

$$x_{d,n+1} = F_x(x_{d,n}, z_{d,n}, \omega_d),$$
$$z_{d,n+1} = F_z(x_{d,n}, z_{d,n}, \omega_d),$$
$$x_{r,n+1} = G_x(x_{r,n}, z_{r,n}, x_{d,n}, z_{d,n}, \omega_d, \varepsilon), \qquad (3)$$
$$z_{r,n+1} = G_z(x_{r,n}, z_{r,n}, x_{d,n}, z_{d,n}, \omega_d, \varepsilon).$$

KORONOVSKIĬ et al.

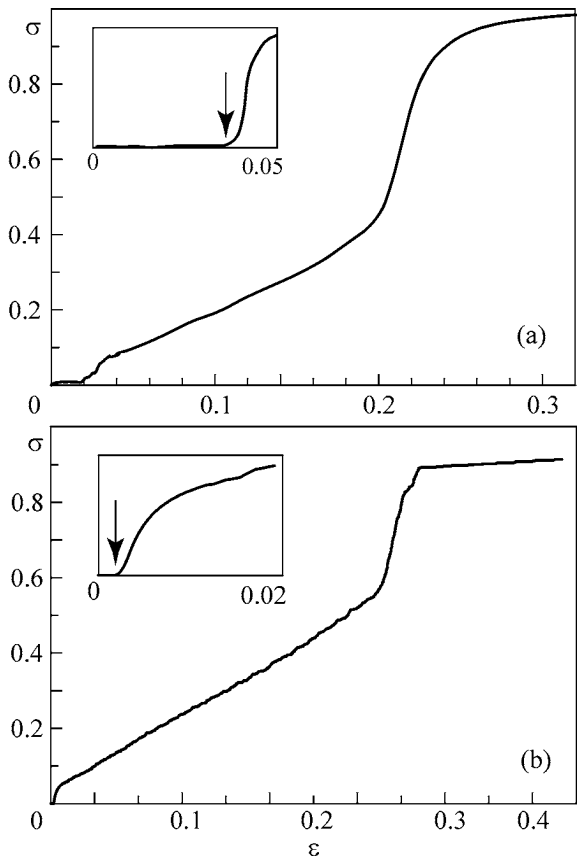

**Fig. 2.** Plots of the geometric measure of synchronism $\sigma(\varepsilon)$ for (a) maps (3) obtained from the Rössler flows (1) by means of the Poincaré sectioning and (b) the logistic maps (2). The insets show the behavior of $\sigma(\varepsilon)$ in the vicinity of zero in a greater scale; arrows indicate the onset of phase synchronization in the Rössler systems $\sigma(\varepsilon)$ and the corresponding point for the coupled maps.

A criterion of synchronism in the coupled maps is provided by the condition that the points in the $(x_n, y_n)$ plane adhere to the diagonal $y_n = x_n$ (where $x_n$ and $y_n$ are the states of the interacting maps at the $n$th moment of discrete time. If the behavior of coupled maps corresponds to the $y = x$ diagonal in the $(x_n, y_n)$ plane, the system features the regime of complete synchronization. If the points are scattered over this plane, the regime is asynchronous.

Let us consider the $(x_{d,n}, x_{r,n})$ plane of maps (3) obtained from flows (1) by means of the Poincaré section for various values of the coupling parameter $\varepsilon$ (Fig. 1a). For the coupling parameter $\varepsilon = 0.015$ at which the phase synchronization is absent (a threshold for the phase synchronization onset with the given control parameters is $\varepsilon_p \approx 0.04$), the points uniformly fill a square in the $(x_{d,n}, x_{r,n})$ plane. If the phase synchronization takes place ($\varepsilon = 0.10$ and $0.225$), the points in the $(x_{d,n}, x_{r,n})$ plane occupy a region having the shape of an irregular quadrangle extended along the $x_r = x_d$ diagonal. The area $S = S(\varepsilon)$ of this region decreases with increasing the coupling parameter, is maximal in the absence of coupling, and tends to zero (the points adhere to the diagonal) in the case of complete synchronization.

Thus, we can introduce the geometric measure defined as

$$\sigma(\varepsilon) = \frac{S(0) - S(\varepsilon)}{S(0)}, \quad (4)$$

which characterizes the degree of synchronization of the interacting systems, where $S(\varepsilon)$ is the area of a region covered by points in the $(x_{d,n}, x_{r,n})$ plane at a fixed value of the coupling parameter $\varepsilon$.

Figure 2a shows the behavior of $\sigma(\varepsilon)$ for maps (3). As can be seen, $\sigma(\varepsilon)$ is close to zero in the absence of phase synchronization, increases with the coupling parameter in the interval corresponding to the phase synchronization, and tends to unity in the complete synchronization regime. It should be noted that the behavior of the proposed geometric measure of synchronism of the interacting systems is consistent with the energetic measure of synchronism introduced previously [10]. Thus, even using the analysis of maps obtained by means of the Poincaré sectioning, it is possible to form a conclusion on the synchronous behavior in a system despite the fact that a significant part of the information about the behavior of the initial flow system was excluded from the consideration.

Now let us consider the behavior of one-way coupled logistic maps (2). In this case, the initial flow system (such as in the above example) is absent and, hence, it is impossible to judge on the presence (or absence) of phase synchronization. Nevertheless, the behavior of coupled systems with discrete time is completely analogous to the behavior of maps (3) obtained by reduction of the Rössler systems (see Figs. 1b and 2b). A figure covered by points in the $(x_n, y_n)$ plane also has the shape of a quadrangle extended along the $x = y$ diagonal, its area decreases with increasing coupling parameter $\varepsilon$, and the geometric measure of synchronism $\sigma(\varepsilon)$ monotonically grows in a certain interval of $\varepsilon$ and tends to unity in the case of complete synchronization. It is also important to note the presence of a certain interval $[0, \varepsilon_p]$ of the coupling parameter where this measure is zero, which corresponds to asynchronous behavior of the maps obtained by reduction of the flow systems (see the insets in Figs. 2a and 2b).

Thus, based on the above considerations, we may conclude that the behavior of the two coupled systems with discrete time observed when the complete synchronization regime is broken as the coupling parameter decreases corresponds to the regime of phase synchronization in slightly non-identical flow systems and has to be considered as a synchronous regime rather than asynchronous as it was commonly accepted until now. Using the proposed quantitative measure of the degree of synchronism in interacting systems, it is possible to unambiguously distinguish the asynchronous behavior of maps from a synchronized regime corresponding to the phase synchronization in coupled flow systems. This approach can be used for the diagnostics of phase synchronism in coupled systems with continuous time.

This study was supported by the Ministry of Education and Science of the Russian Federation (program "Development of Scientific Potential of High School," project no. 333); the Russian Foundation for Basic Research (project no. 05-02-16273); the Council of the President of the Russian Federation for Support of Young Russian Scientists and Leading Scientific Schools (project no. NSh-1250.2003.02); and the Science and Education Center "Nonlinear Dynamics and Biophysics" at the Saratov State University (sponsored by the US Civilian Research and Development Foundation, grant no. REC-006). We also gratefully acknowledge support from the "Dynasty" Foundation and the International Center for Fundamental Physics in Moscow (ICFPM).

*Translated by P. Pozdeev*